\documentstyle[twocolumn,aps]{revtex}
\input epsf
\begin{document}
\draft
\twocolumn[\hsize\textwidth\columnwidth\hsize\csname@twocolumnfalse\endcsname

\title{Anomalous Spreading of Power-Law Quantum Wave Packets}
\author{Fabrizio Lillo and Rosario N. Mantegna}
\address{
Istituto Nazionale per la Fisica della Materia, Unit\`a di Palermo\\
and\\
Dipartimento di Energetica ed Applicazioni di Fisica,
Universit\`a di Palermo, Viale delle Scienze, I-90128,
Palermo, Italia}
\maketitle
\begin{abstract}

We introduce power-law tail quantum wave packets. We show that they can
be
seen as eigenfunctions of a Hamiltonian with a physical potential.
We prove that the free evolution of these packets presents an asymptotic 
decay of the maximum of the wave packets which is anomalous for an
interval 
of the characterizing power-law exponent. We also prove that the number
of 
finite moments of the wave packets is a conserved quantity during the 
evolution of the wave packet in the free space.

\end{abstract}
\vskip2pc]

\narrowtext

Power-law probability density functions \cite{Levy25} are receiving a 
lot of attention in different research fields 
\cite{Mandelbrot82,Refpl}. 
Stochastic processes with power-law distributions may or may not be 
characterized by a typical scale in time and/or in the size of the 
random variable. One implication of the absence of a typical scale 
is the divergence of the variance of the distribution. Examples of 
phenomena without a typical scale are observed in physical systems 
at the critical state \cite{Stanley}, in self-organized \cite{Bak} 
and in complex systems \cite{Complex}. Considering few-body quantum 
systems - (i) scale free power-law processes have been observed in a 
quantum system by investigating and modeling experiments of velocity 
selective coherent population trapping \cite{Bardou94,Reichel95} 
and (ii) the power-law temporal growth of the moments of a wave packet 
has been investigated in quantum systems with fractal energy spectra and 
eigenfunctions such as the Harper model \cite{Geisel91,Ketzmerick97}.
Spatial power-law wave functions have not been considered within the
framework 
of non-relativistic quantum mechanics. The probabilistic interpretation 
of the wave function and the recent results on power-law distributions 
observed in physical systems motivates us to investigate the properties 
of power-law wave functions in quantum mechanics. 
In this letter we consider quantum wave packets with power-law tails. 
Specifically we focus on -- (i) their physical properties. Namely the 
uncertainty product, the associated energy and the momentum distribution 
and (ii) the spreading of such quantum wave packets during the free
evolution.

We define as Power-Law Tail Wave Packet (PLTWP) a wave function $\psi
(x)$ 
describing a non-relativistic spinless particle in one dimension, 
which decreases with $x$ as
 
\begin{equation}
\mid\psi(x)\mid\sim\mid x\mid^{-\alpha}. 
\end{equation}

\noindent
This class of wave packets is square-integrable only if $\alpha > 1/2$.
For the sake of simplicity, in this study we assume that wave function
is 
real, positive and even. The study of quantum wave packets with wave 
function real or complex, uneven and with zeros is presented elsewhere
\cite{Lillo99}.
One of the properties of power-law distributions is that only a finite
number 
of moments of the variable are finite. In quantum theory this implies
that the 
moments of position operator $\langle \hat x^m \rangle$ with $m \geq
2\alpha-1$ 
are infinite. The lack of finite moments of $ x$ has a counterpart in
the 
properties in $k=0$ of Fourier transform $g(k)=FT[\psi(x)]$ which gives
the 
amplitude probability distribution of momentum. 
We note that when $n<\alpha \leq n+1$, only the first $n-1$ derivatives
of $g(k)$ 
exist in $k=0$. In the special case $1/2 <\alpha \leq 1$ the Fourier
transform $g(k)$  
is infinite in $k=0$. The behavior of $g(k)$ when $k \simeq 0$ is 
described by two different series expansions depending on the value of
$\alpha$.
Specifically, when $\alpha$ is not an odd integer number 
\begin{equation}
g(k)\simeq a_0+a_2 k^2+...+a_{2n}k^{2n}+b\mid k\mid ^{\alpha-1}+o(\mid k
\mid ^{\alpha}),
\end{equation}
where $2n$ is the largest even integer number smaller than $\alpha-1$.
When $\alpha = 2n+1$ the series expansion is
\begin{equation}
g(k)\simeq a_0+a_2 k^2+...+a_{2n}k^{2n}+b k^{2n}\log \mid k \mid +...~ .
\end{equation}
The case $\alpha =1$ has a logarithmic divergence in $k=0$.
  
In spite of this behavior we wish to point out that all moments of the 
momentum operator (including the kinetic energy) of the particle 
$\langle \hat p^m \rangle$ ($m \geq 1$) are finite. 
In fact, a property of Fourier transform is

\begin{equation}
FT[\psi^{(m)}(x)]=(-ik)^m g(k), 
\end{equation}
where $\psi^{(m)}(x)$ indicates the $m$-th derivative of $\psi(x)$. This
property 
is valid under the hypothesis that 
$\lim_{|x| \rightarrow \infty} \psi^{(r)}(x)=0$ for $r=0,1,...,m-1$.  
Since the tails behave as $\mid x \mid ^{-\alpha -m}$, $\psi^{(m)}(x)$
is 
absolutely integrable. By using Riemann-Lebesgue lemma
\cite{Lighthill70}, 
one can conclude that $\lim_{k \rightarrow \infty} k^m g(k)=0$ for all
$m$. 

A consequence of the finiteness of momentum moments is that the
uncertainty 
in momentum $\Delta k$ is always finite for PLTWP. As mentioned above, 
when $\alpha \leq \frac{3}{2}$ the second moment of the position
operator and 
thus the root mean square deviation $\Delta x$ of position are infinite. 
Therefore this kind of wave packets have an uncertainty product 
$\Delta x \Delta k$ which is infinite when $\alpha \leq \frac{3}{2}$.   
This physical property reflects the fact that for PLTWP with 
$\alpha \leq \frac{3}{2}$ a typical scale exists in momentum space
whereas 
the wave packet is scale free in space.

Can a PLTWP be an eigenfunction 
of an Hamiltonian? To answer this
question we 
note that each PLTWP can be related to a specific physical potential.
The PLTWP is then an eigenfunction of the corresponding Hamiltonian. The
equation 
providing the potential associated with a particle described by a wave
function 
$\psi(x)$ is \cite{Lamb69}
\begin{equation}
U(x)= E + \frac{\hbar ^2}{2M} \frac{1}{\psi(x)}\frac{d^2\psi(x)}{dx^2},  
\end{equation}
where $E$ is the eigenvalue of energy and $M$ is the mass of the
particle.
The shape of the potential depends on the local properties of the wave
function. 
>From Eqs. (1) and (5) it is immediate to conclude that the potential
associated with 
PLTWP behaves asymptotically as $x^{-2}$ \cite{note1}. As an
illustrative example we 
present PLTWPs defined as 
\begin{equation}
\psi(x)=\frac{N}{(x^2+\gamma ^2)^{\alpha/2}},
\end{equation}
where $N$ is a suitable normalization constant and $\gamma$ is a scale
parameter. 
The above family of quantum wave packets is related to the Student's 
{\it t}-distribution when $\alpha$ is integer. The associated potential
is 
\begin{equation}
U(x)=\frac{\hbar^2}{2M}\alpha
\frac{x^2(1+\alpha)-\gamma^2}{(x^2+\gamma^2)^2},
\end{equation}
and is shown in Fig. 1. In this 
figure the eigenvalue $E$ is set equal to zero. The potential is a
single well 
potential with two symmetrical confining barriers. The potential reaches
a maximal 
value and then decreases asymptotically as $U(x) \sim x^{-2}$. A general
property 
of this potential is that by increasing the value of $\alpha$ the depth
of the 
potential well increases. From Fig. 1 it is clear that the associated
potential 
does not present anomalies of any sort.

Hereafter we consider the properties of a PLTWP in the simplest case 
of dynamical evolution, namely the free wave evolution. We assume that 
a $t=0$ the wave packet in the free space has the asymptotic properties
of 
Eq. (1). We focus our attention on the spreading of the wave packet. 
During the dynamical evolution in free space the wave function at time
$t$ 
is given by
\begin{equation}
\psi(x,t)=\frac{1}{\sqrt{2\pi}} \int_{-\infty}^{\infty}
g(k)e^{i(kx-\hbar tk^2/2M)} dk .
\end{equation}
We briefly recall that in the case of a Gaussian wave packet the amount 
of spreading of the wave packet can be quantified either by considering 
the time dependence of the position variance or by determining the time 
dependence of the maximum of the wave function.
To quantify the amount of spreading of a free wave packet in the same 
way for PLTWP with finite or infinite variance we choose to focus our 
attention to the asymptotic behavior in time of the wave function in a 
specific position (for example $x=0$). 
In the Gaussian case, the variance of $\mid \psi (x,t) \mid ^2$ is
asymptotically 
proportional to $t^2$ and the maximum of the wave function $\mid
\psi(0,t) 
\mid ^2$ decreases as $t^{-1}$ asymptotically. We will show in the
following 
that this asymptotic behavior observed for Gaussian quantum packets is
not 
universally observed in the free evolution of a quantum wave packet.  

We consider here the free evolution of the maximum of the packet, which
is 
described by Eq. (8) with $x=0$. It is possible to give an asymptotic 
expansion in time $t$ of integral of Eq. (8) using phase stationary 
method (see for example \cite{Olver74}). This method shows that the
asymptotic 
expansion of $\psi(0,t)$ is determined by the dominant term of series
expansion 
about the origin ($k=0$) of $g(k)$ and by the function $-\hbar t
k^2/2M$.
The first term of the series expansion about $k=0$ of $g(k)$ is 
$g(k)\simeq Q |k|^{\lambda -1}$ for any value of $\alpha \neq 1$. 
By using phase stationary method \cite{Olver74}, we know that when 
$0< \lambda < 2$ it is possible to give an asymptotic expansion of
integral 
of  Eq. (8) with $x=0$ in the form
\begin{equation}
\psi (0,t) \sim \frac{1}{\sqrt{2\pi}} e^{-i\pi \lambda/4} Q 
\Gamma (\frac{\lambda}{2})\frac{1}{\beta^{\lambda/2}}, 
\end{equation}   
where $\beta=\hbar t/2M$. If $g(k)$ has a finite non-vanishing limit 
as $k$ goes to zero then $\lambda=1$ and we conclude that the maximum 
of the packet decreases asymptotically in time as $1/\sqrt{t}$. This 
is the case, for example, of Gaussian wave packets cited above and, in
general, 
of all commonest wave packets with finite FT in $k=0$. From Eqs. (2,3)
and (9) 
we conclude that this is also the case of PLTWP with $\alpha >1$. These
packets 
show a customary asymptotic dynamics of the maximum. PLTWPs with 
$1/2<\alpha <1$ show a different behavior. In fact, in these cases
$g(k)$ diverges in $k=0$ as $\mid k \mid ^{\alpha-1}$ and, as a consequence 
$\lambda =\alpha < 1$.
 Hence, form Eq. (9) we conclude that the maximum
decreases 
asymptotically as $t^{-\alpha/2}$. This is a maximum decrease, 
which is {\it anomalous} and slower with respect to the decrease
observed in 
customary wave packets. The $t^{-1/2}$ behavior observed in Gaussian
wave 
packets is interpreted in terms of the time evolution of a group of
classical 
particles with a momentum dispersion $\Delta p$. A similar simple
picture 
cannot explain the behavior observed for PLTWPs with $\alpha<1$
\cite{note2}.

To illustrate the process of convergence of different wave packets to
the 
expected asymptotic behavior we calculate the time evolution of the 
amplitude of wave packet at $x=0$ for three different cases.
Specifically 
we consider a Gaussian wave packet and two PLTWPs of the class defined
by 
Eq. (6) with $\alpha=3$ and $\alpha=0.75$. In Fig. 2 we show the
numerical 
estimate of $|\psi(0,t)|$. In the figure is clear that the Gaussian and 
the PLTWP with $\alpha=3$ soon converge to the usual $1/\sqrt{t}$
asymptotic 
behavior whereas the PLTWP with $\alpha=0.75$ slowly converges to the 
anomalous asymptotic behavior of $1/t^{\alpha/2}=1/t^{0.375}$.
 
The case $\alpha =1$ cannot be handled with phase stationary method 
because of logarithmic divergence of $g(k)$ in $k=0$. Although we are
not 
able to provide a general answer for the case $\alpha=1$, we are able to 
determine the asymptotic behavior in the specific case of a packet
described 
by Eq. (6) with $\alpha=1$. For a such wave packet, the maximum
decreases 
as $1/\sqrt{t}$. 

In order to obtain a more complete description of the free wave
spreading, 
we consider the time evolution of the tails of the packet. We prove that 
the free wave packet evolution of a PLTWP conserves the tails. More
precisely, 
if at $t=0$ the dominant term of asymptotic expansion of $\psi(x)$ is 
$c \mid x \mid ^{-\alpha}$, at each subsequent time the asymptotic
expansion 
of $\psi(x,t)$ will be dominated by $c \mid x \mid ^{-\alpha}$. In this
sense, 
the free evolution cannot change the asymptotic properties of the
packet. 
We can visualize this property dividing the whole set of PLTWPs in
classes 
such as each class is characterized by a value of $\alpha$. In other
words, 
any packet belongs to the same class at any time during the free wave
evolution. 
Since $\alpha$ determines the number of finite moments of operator $\hat
x$, 
our result implies that position moments cannot became finite if they
were 
infinite at $t=0$ and vice versa. 
We prove our statement as follows. Let us first consider Eq. (8) and
look 
for the asymptotic expansion in $x$ of the FT of $f(k)=g(k)e^{-i\beta
k^2}$. 
The asymptotic expansion of the FT of $f(k)$ is determined by its
singularities 
\cite{Lighthill70}. By following Ref. \cite{Lighthill70}, we say that
$f(k)$ 
is singular in $k_0$ if one cannot differentiate $f(k)$ in $k_0$ any
number of 
times. From properties of $g(k)$ of PLTWPs we conclude that $k=0$ is the
only 
singularity for $f(k)$. In order to find the asymptotic expansion of Eq.
(8), 
we construct a function $F(k)$ in such a way that $\tilde f(k)\equiv
f(k)-F(k)$ 
has absolutely integrable $m$-th derivative in an interval including
$k=0$. $F(k)$ 
must be a linear combination of powers of $k$ and product of powers of
$k$ and 
$\log k$ \cite{Lighthill70}. Moreover the $m$-th derivative of $f(k)$
must be 
absolutely integrable in an interval from a real value to infinity. If
all 
these hypothesis are verified, the FT of $f(k)$ is equal to the FT of
$F(k)$ 
plus $o(\mid x \mid ^{-m})$. Depending on the specific value of $\alpha$
the 
regularizing function assumes a different form. In spite of this, our
results 
do not depend on the specific value of $\alpha$. For the sake of
simplicity, 
we present here the case with $2n+1<\alpha<2n+2$. The demonstration of
cases 
for other value of $\alpha$ (specifically $2n<\alpha<2n+1$, $\alpha=2n$
and 
$\alpha=2n+1$) are similar. In order to regularize the $2n+1$-th
derivative 
of $f(k)$ the natural choice is
\begin{equation}
\tilde f(k)=g(k)e^{-i\beta k^2}-b\mid k \mid ^{\alpha -1} .
\end{equation}
In fact the first $2n+3$ derivatives of $\tilde f(k)$ are absolutely
integrable 
in an interval including $k=0$. Therefore, we obtain the asymptotic
expansion of 
$\psi (x,t)$ as
\begin{eqnarray}
\psi (x,t)=FT(F(k))+o(\mid x\mid ^{-2n-3})= \nonumber \\
c~\mid x \mid ^{-\alpha}+ o(\mid x\mid^{-2n-3})
\end{eqnarray}
which demonstrates our assertion. 

As an illustrative example let us consider the free evolution of the
PLTWP of 
Eq. (6) with $\alpha =2$ and $\gamma =1$. The corresponding wave
function has 
the form of the Cauchy distribution and it is possible to find the $\psi
(x,t)$ 
analytically. In Fig. 3 we show the square modulus of the positive tail
of the 
packet versus $x$ at different times in a log-log plot. The figure shows
that 
the dominant term of the asymptotic expansion, i.e. $x ^{-4}$, is the same 
at any time. The figure also shows that new terms of the
asymptotic expansion become relevant at longer times    
and the region of asymptotic
convergence 
moves towards larger values of $x$.    

A connection between stochastic processes and quantum processes has been 
considered within the framework of stochastic mechanics \cite{Guerra81}.
Here we note that the free evolution of a quantum wave packet is closely 
related to a superdiffusive (ballistic) stochastic process.
When a superdiffusive behavior is observed in classical 
\cite{Cassandro78,Shlesinger93,Zaslavsky97} and quantum
\cite{Sundaram99} processes, 
the central limit theorem does not apply and a general 
theoretical description is lacking. In our study we obtain two general
conclusions 
in the problem of the free evolution of a PLTWP. The first concerns the
temporal 
evolution of the maximum of the wave packet (which corresponds to the 
probability of return to the origin in stochastic processes). In the
well-known 
case of a Gaussian wave packet (which has a fractional Brownian motion
with 
exponent $h=1$ as the approximately corresponding stochastic process 
\cite{Mandelbrot68}) the variance is quadratic in time and the
probability of 
return to the origin is inversely proportional to the time. 
Similarly, we find that the same behavior is asymptotically observed for
the 
evolution of an even and real PLTWP when the exponent $\alpha$ is
greater than 
one. Conversely, for values of $\alpha$ within the interval
$1/2<\alpha<1$ we 
observe an {\it anomalous} behavior of the time evolution of the maximum
of 
the wave packet. 
We are not able to interpret this unexpected behavior on a semiclassical
basis. 
The second conclusion concerns the conservation during the quantum time 
evolution of the number of finite/infinite moments of the $t=0$
distribution. 
This behavior is peculiar to this quantum dynamics and could not be
observed, 
for example, in stochastic processes obeying the central limit theorem.

We thank INFM and MURST for financial support. We wish to thank Giovanni 
Bonanno for help in numerical calculations.


\begin{figure}[t]
\epsfxsize=2.6in
\epsfbox{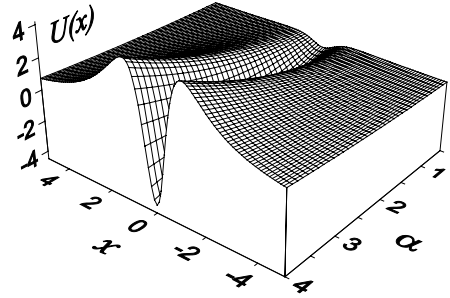}
\caption{Potential profile, Eq. (7), associated with 
the class of power-law tail wave 
packets of Eq. (6).  Here we show a region of $\alpha$ values ranging
from 
the minimal allowed value $\alpha=1/2$ (excluded) to the value
$\alpha=4$. 
In the entire interval $\alpha>1/2$, the potential does not show any 
pathological behavior.
In our calculation we set $\gamma=1$. The potential $U(x)$ is given in
units 
of $\hbar^2/2M$. The PLTWP is the eigenfunction of the corresponding 
Hamiltonian with eigenvalue $E=0$.}
\label{fig1}
 \end{figure}
 
 \begin{figure}[t]
\epsfxsize=2.6in
\epsfbox{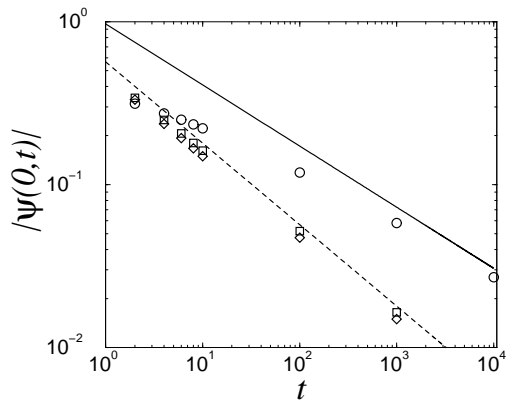}
\caption{Log-log plot of the time evolution of the wave packet amplitude
at 
$x=0$ for three different wave packets. The usual asymptotic behavior
$t^{-1/2}$ 
is observed for the Gaussian (diamond) and for the PLTWP defined by Eq.
(6) 
with $\alpha=3$ (square). The anomalous decay predicted for PLTWP with 
$1/2<\alpha<1$ is observed for the PLTWP defined by Eq. (6) with
$\alpha=0.75$ 
(circle). To estimate the degree of convergence, we also show the
asymptotic 
behaviors predicted in the two cases as straight lines. The dashed line 
indicates the $t^{-1/2}$ behavior and the solid line shows the
$t^{-0.375}$ 
behavior. In our calculation we set $\gamma=1$, $\hbar/2M=1$ and we
chose 
the value of $\Delta x$ of the Gaussian wave packet equal to the same
quantity 
of the PLTWP with $\alpha=3$ at $t=0$.}
\label{fig2}
 \end{figure}

\begin{figure}[t]
\epsfxsize=2.6in
\epsfbox{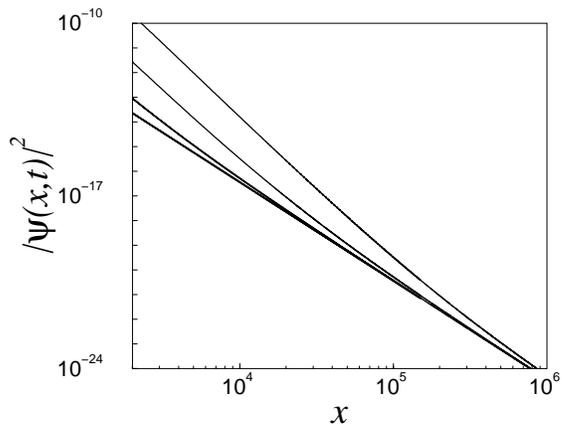}
\caption{Log-log plot of the positive tail of $|\psi(x,t)|^2$ for the PLTWP of 
Eq. (6) with $\alpha=2$ as a function of $x$ for different values of the
time. 
The bottom line refers to the case $t=0$. From bottom to top the other
lines 
describe the cases $t=10^4$, $t=10^5$, and $t=10^6$ respectively. The
dominance 
of the leading term of the asymptotic expansion at $t=0$ is also
observed for 
longer times at larger values of $x$. In our calculation we set
$\gamma=1$ 
and $\hbar/2M=1$.}
\label{fig3}
 \end{figure}


\begin{references}
\bibitem{note1}  The power-law behavior of $\psi(x)$ is also 
observed for large values of $|x|$ for eigenfunctions with energy $E^*$ 
greater than $E$ provided that $|x|\ll\sqrt{\hbar^2 \alpha (\alpha+1)/
2 M (E^*-E)}$. The existence of bound states and their properties 
depends on the specific form of $U(x)$ nearby $x=0$.
\bibitem{note2}  The analogy 
between the Wigner function and the corresponding classical 
probability distribution in the phase space cannot be used for 
PLTWPs because the Wigner function of such quantum states 
becomes negative in some regions of the phase space 
(for a proof see Ref. [11]). Hence a simple analogy cannot
be pursued in spite of the fact that the 
dynamical equation for the Wigner function is identical with 
the classical Liouville equation in the free evolution.
\end{references}
\end{document}